\documentclass{ws-ijgmmp}
\usepackage{amsmath,amssymb}

\renewcommand{\vec}[1]{\boldsymbol{\mathrm{#1}}}
\DeclareMathOperator{\Tr}{Tr}

\makeatletter
\newcommand{\verbatimfont}[1]{\renewcommand{\verbatim@font}{\ttfamily#1}}
\makeatother

\begin{document}

\title{Efficient trace-free decomposition of symmetric tensors of arbitrary rank}

\author{Viktor T. Toth}
\address{Ottawa, Ontario K1N 9H5, Canada}
\author{Slava G. Turyshev}
\address{Jet Propulsion Laboratory, California Institute of Technology,\\
4800 Oak Grove Drive, Pasadena, CA 91109-0899, USA}

\maketitle

\begin{abstract}
Symmetric trace-free tensors are used in many areas of physics, including electromagnetism, relativistic celestial mechanics and geodesy, as well as in the study of gravitational radiation and gravitational lensing. Their use allows integration of the relevant wave propagation equations to arbitrary order. We present an improved iterative method for the trace-free decomposition of symmetric tensors of arbitrary rank. The method can be used both in coordinate-free symbolic derivations using a computer algebra system and in numerical modeling. We obtain a closed-form representation of the trace-free decomposition in arbitrary dimensions. To demonstrate the results, we compute the coordinate combinations representing the symmetric trace-free (STF) mass multipole moments for rank 5 through 8, not readily available in the literature.
\end{abstract}

\keywords{STF tensors,trace-free decomposition,computer algebra}

\ccode{Mathematics Subject Classification 2020: 15A72, 68W30, 78A10, 85-04}

\section{Introduction}

In our investigation of gravitational lensing phenomena \cite{SGL2021a,SGL2021f,SGL2021g}, we utilized the representation of  the scalar gravitational potential via an infinite series of symmetric trace-free (STF) tensor mass multipole moments \cite{Hamermesh1962}.
The STF mass multipole moments of an object of arbitrary shape can be obtained by integrating the appropriate STF coordinate combinations over the volume of the object, as was demonstrated in \cite{SGL2021g}. Such a representation is useful to study not just gravitational lensing but also electromagnetism, gravitational radiation, geodesy and celestial mechanics \cite{1974AIHPA..21..319Y,1980RvMP...52..299T,1994CeMDA..60..139H,2014IJMPD..2350003Z,2017JGeod..91..783S,Soffel-Han:2019}.

Although the STF representation of a scalar potential is equivalent to the representation using spherical harmonics, it simplifies the integration of wave propagation equations \cite{SGL2021a,SGL2021f,SGL2021g} and offers the advantage of coordinate independence \cite{2016CQGra..33a5008C}. This, in turn, makes it possible to use a Cartesian coordinate system for actual calculations, which offers great convenience over spherical coordinates in many scenarios. On the other hand, the STF formalism can be computationally intensive, which may explain why it remains somewhat less well known and less frequently used than spherical harmonics.

As we shall see below, the trace-free representation of tensors is unique. Obtaining the decomposition, however, requires solving a potentially very large linear system of equations \cite{Hamermesh1962}. The complexity of the system of equations makes it difficult to use the results in analytical derivations. While it is computationally straightforward to solve a linear system of equations numerically, such solutions are opaque, offer little insight, and are useless when the goal is to obtain closed form analytical or semianalytical solutions of a physical problem.

On the other hand, the problem can be greatly simplified when the tensor in question is symmetric in all indices \cite{1968JMP.....9....1T}. This is the case in particular when it comes to tensorial STF mass multipole moments. This investigation led us to a new, direct, coordinate-independent method of trace-free decomposition of fully symmetric tensors of arbitrary rank. After implementing this method using computer algebra software, we realized that a more efficient, closed-form representation is also achievable. Guided by earlier efforts to obtain such a representation specifically in the three-dimensional case, we were able to derive a closed-form result in arbitrary dimensions.

In the present paper, we offer a brief introduction to the trace-free representation in Sec.~\ref{sec:tracefree}. In Section~\ref{sec:symmetric} we explore the fully symmetric case. In Section~\ref{sec:closedform} we present a closed-form representation for the STF coordinate combinations. We offer a brief discussion in Sec.~\ref{sec:conclusions}. We demonstrate the utility of this result in the Appendix by way of a simple use case, computing the coordinate combinations that represent STF mass multipole moments for rank 5 through 8.

\section{Trace-free representation of a generic tensor}
\label{sec:tracefree}

To introduce our approach, let us first consider the simplest case, the trace of a rank-2 tensor:
\begin{align}
\Tr\mathbb{Q}=Q^a_a=Q=\eta_{ab}Q^{ab},
\end{align}
where $\eta_{ab}$ is the covariant metric. (In case of Euclidean space with Cartesian coordinates, $\eta_{ab}=\delta_{ab}$, but our approach is not confined to this special case; we assume a generic metric manifold with a nondegenerate metric.) The trace-free representation amounts to decomposing the generic tensor $Q^{ab}$ into a linear combination of a trace-free tensor $Q_0^{ab}$ and the trace $Q$ multiplied by  the metric:
\begin{align}
Q^{ab}=Q_0^{ab}+\frac{1}{n}Q\eta^{ab},
\label{eq:Q0-r2}
\end{align}
where $n$ is the number of dimensions. It is well known and easy to see that this representation is unique.

For rank-3 tensors the decomposition becomes a bit more interesting. These tensors have three distinct traces, $\eta_{ab}Q^{abc}$, $\eta_{bc}Q^{abc}$ and $\eta_{ca}Q^{abc}$. The decomposition then follows:
\begin{align}Q^{abc}=Q_0^{abc}+{\cal A}^a\eta^{bc}+{\cal B}^b\eta^{ca}+{\cal C}^c\eta^{ab}.\label{eq:Tr3}
\end{align}
To determine the values of the quantities ${\cal A}^a$ through ${\cal C}^a$, we can take the trace of (\ref{eq:Tr3}) and obtain three linear tensor equations in three sets of unknowns:
\begin{align}
\eta_{bc}Q^{abc}&=n{\cal A}^a+{\cal B}^a+{\cal C}^a,\\
\eta_{ca}Q^{abc}&={\cal A}^b+n{\cal B}^b+{\cal C}^b,\\
\eta_{ab}Q^{abc}&={\cal A}^c+{\cal B}^c+n{\cal C}^c.
\end{align}
These can be readily solved to yield
\begin{align}
{\cal A}^i&=\frac{1}{(n-1)(n+2)}\left[(n+1)\eta_{jk}Q^{ijk}-\eta_{jk}Q^{kij}-\eta_{jk}Q^{jki}\right],\\
{\cal B}^i&=\frac{1}{(n-1)(n+2)}\left[-\eta_{jk}Q^{ijk}+(n+1)\eta_{jk}Q^{kij}-\eta_{jk}Q^{jki}\right],\\
{\cal C}^i&=\frac{1}{(n-1)(n+2)}\left[\eta_{jk}Q^{ijk}-\eta_{jk}Q^{kij}+(n+1)\eta_{jk}Q^{jki}\right],
\end{align}
thus
\begin{align}
Q_0^{abc}=Q^{abc}-{\cal A}^a\eta^{bc}-{\cal B}^b\eta^{ca}-{\cal C}^c\eta^{ab}.
\end{align}

This establishes the generic pattern. As we can see, the representation is unique. For rank-$k$ tensors, the representation involves $\tfrac{1}{2}k(k-1)$ unknown tensors of rank $(k-2)$:
\begin{align}
Q^{ab..k}=Q_0^{ab..k}+{\cal A}^{c..k}\eta^{ab}+{\cal B}^{bd..k}\eta^{ac}+...+{\cal X}^{ab..i}\eta^{jk},\label{eq:main}
\end{align}
which, after taking the traces, yields $\tfrac{1}{2}k(k-1)$ equations in the form
\begin{align}
\eta_{ab}Q^{ab..k}&=n{\cal A}^{c..k}+{\cal B}^{c..k}+...+\eta_{ab}{\cal X}^{ab..i}\eta^{jk},\\
...\nonumber\\
\eta_{jk}Q^{ab..k}&=\eta_{jk}{\cal A}^{c..k}\eta^{ab}+\eta_{jk}{\cal B}^{bd..k}\eta^{ac}+...+n{\cal X}^{ab..i}.
\end{align}
The structure of this system of equations is more complex than it appears at first sight, as it sometimes involves the unknown tensors, sometimes their traces. In the general case, this is a linear system of equations in $\tfrac{1}{2}k(k-1)n^2$ unknowns. In principle, this is trivially solvable, but the result is a complex expression, not readily tractable.

\section{Symmetric trace-free tensors}
\label{sec:symmetric}

When $Q^{ab..k}=Q^{(ab..k)}$ is symmetric in all indices, the situation is greatly simplified, since ${\cal A}={\cal B}=...={\cal X}$ in (\ref{eq:main}). In particular, for $k=3$ we get
\begin{align}
Q^{abc}=Q_0^{abc}+\eta^{ab}{\cal A}^{c}+\eta^{ac}{\cal A}^{b}+\eta^{bc}{\cal A}^{a}.\label{eq:Qabc}
\end{align}
Taking the trace with respect to the $ab$ indices, we get
\begin{align}
\eta_{ab}Q^{abc}=Q^{c}=(n+2){\cal A}^{c}.
\label{eq:Qc}
\end{align}
Solving for ${\cal A}^c$ and substituting back into (\ref{eq:Qabc}), we obtain
\begin{align}
Q_0^{abc}=Q^{abc}-\frac{1}{n+2}\left(\eta^{ab}{ Q}^{c}+\eta^{ac}{ Q}^{b}+\eta^{bc}{ Q}^{a}\right).
\label{eq:Q0-r3}
\end{align}
Similarly, for $k=4$, we obtain the equation
\begin{align}
Q^{abcd}=Q_0^{abcd}+\eta^{ab}{\cal A}^{cd}+\eta^{ac}{\cal A}^{bd}+\eta^{ad}{\cal A}^{bc}+\eta^{bc}{\cal A}^{ad}+\eta^{bd}{\cal A}^{ac}+\eta^{cd}{\cal A}^{ab}.\label{eq:Qabcd}
\end{align}
Taking the trace with respect to the $ab$ indices, we get
\begin{align}
\eta_{ab}Q^{abcd}=Q^{cd}=(n+4){\cal A}^{cd}+\eta^{cd}{\cal A}.\label{eq:Qcd}
\end{align}
Taking the trace again, with respect to $cd$, we obtain
\begin{align}
{\cal A}=\frac{1}{2n+4}Q,
\end{align}
which we can substitute back into (\ref{eq:Qcd}), to get
\begin{align}
{\cal A}^{cd}=\frac{1}{n+4}\left(Q^{cd}-\eta^{cd}\frac{1}{2n+4}Q\right).\label{eq:Acd}
\end{align}
Finally, we can substitute this result back into (\ref{eq:Qabcd}):
\begin{align}
Q_0^{abcd}=Q^{abcd}-\frac{6}{n+4}\eta^{(ab}\left(Q^{cd)} - \frac{1}{2n+4}\eta^{cd)}Q\right),
\label{eq:Q0-r4}
\end{align}
where we used round brackets in indices to indicate symmetrization. This nice, clean result suggests that a generalization to arbitrary rank $k$ might be possible. To investigate this, let us first consider the $k=5$ case following the same approach:
\begin{align}
Q^{abcde}&{}=Q_0^{abcde}+\eta^{ab}{\cal A}^{cde}+\eta^{ac}{\cal A}^{bde}+\eta^{ad}{\cal A}^{bce}+\eta^{ae}{\cal A}^{bcd}+\eta^{bc}{\cal A}^{ade}\nonumber\\
&\phantom{{}=Q_0^{abcde}}+\eta^{bd}{\cal A}^{ace}+\eta^{be}{\cal A}^{acd}+\eta^{cd}{\cal A}^{abe}+\eta^{ce}{\cal A}^{abd}+\eta^{de}{\cal A}^{abc},\\
\eta_{ab}Q^{abcde}=Q^{cde}&{}=(n+6){\cal A}^{cde}+\eta^{cd}{\cal A}^e+\eta^{ce}{\cal A}^d+\eta^{de}{\cal A}^c,\\
{\cal A}^e&{}=\frac{1}{2n+8}Q^e,\\
{\cal A}^{cde}&{}=\frac{1}{n+6}\left(Q^{cde} - \frac{3}{2n+8}\eta^{(cd}Q^{e)}\right),\\
Q_0^{abcde}&{}=Q^{abcde}-\frac{10}{n+6}\eta^{(ab}\left(Q^{cde)} - \frac{3}{2n+8}\eta^{cd}Q^{e)}\right).
\label{eq:Q0-r5}
\end{align}
However, generalizing to $k=6$ introduces a new twist, the need to take the trace not twice but three times:
\begin{align}
Q^{abcdef}&{}=Q_0^{abcdef}+\eta^{ab}{\cal A}^{cdef}+\eta^{ac}{\cal A}^{bdef}+...+\eta^{ef}{\cal A}^{abcd},\\
\eta_{ab}Q^{abcdef}=Q^{cdef}&{}=(n+8){\cal A}^{cdef}+\eta^{cd}{\cal A}^{ef}+\eta^{ce}{\cal A}^{df}+\eta^{cf}{\cal A}^{de}\nonumber\\
&{}\hskip 1in +\eta^{de}{\cal A}^{cf}+\eta^{df}{\cal A}^{ce}+\eta^{ef}{\cal A}^{cd},\\
\eta_{ab}\eta_{cd}Q^{abcdef}=Q^{ef}&{}=(2n+12){\cal A}^{ef}+\eta^{ef}{\cal A},
\end{align}\begin{align}
{\cal A}&{}=\frac{1}{3n+12}Q,\\
{\cal A}^{ef}&{}=\frac{1}{2n+12}\left(Q^{ef}-\frac{1}{3n+12}\eta^{ef}Q\right),\\
{\cal A}^{cdef}&{}=\frac{1}{n+8}\left[Q^{cdef}-\frac{6}{2n+12}\eta^{(cd}\left(Q^{ef)}-\frac{1}{3n+12}\eta^{ef)}Q\right)\right],\\
Q_0^{abcdef}&{}=Q^{abcdef}-\frac{15}{n+8}\eta^{(ab}\left[Q^{cdef)}-\frac{6}{2n+12}\eta^{cd}\left(Q^{ef)}-\frac{1}{3n+12}\eta^{ef)}Q\right)\right].
\label{eq:Q0-r6}
\end{align}

Even as, with higher rank, we need to calculate more and more traces, this establishes a general pattern that allows us to compute the trace-free representation of fully symmetric tensors:

\begin{align}
Q^{ab..k}&{}=Q_0^{ab..k}+\underbrace{\eta^{ab}{\cal A}^{cd..k}+\eta^{ac}{\cal A}^{bd..k}+...+\eta^{jk}{\cal A}^{ab..i}}_{k(k-1)},\\
Q^{cd..k}&{}=[n+2(k-2)]{\cal A}^{cd..k}+\underbrace{\eta^{cd}{\cal A}^{ef..k}+\eta^{ce}{\cal A}^{df..k}+..+\eta^{jk}{\cal A}^{cd..i}}_{(k-2)(k-3)},\\
Q^{ef..k}&{}=[2n+2(k-2)+2(k-4)]{\cal A}^{ef..k}+\underbrace{\eta^{ef}{\cal A}^{gh..k}+\eta^{eg}{\cal A}^{fh..k}+..+\eta^{jk}{\cal A}^{ef..i}}_{(k-4)(k-5)},\\
...\nonumber
\end{align}

This approach can be readily continued to arbitrary rank, as required.

\section{Coordinate-free decomposition algorithm}
\label{sec:algorithm}

The generic procedure to obtain the trace-free representation of a symmetric tensor can be summarized as follows:

\begin{enumerate}
\item Start with the equation
\begin{align}
Q^{ab..k}=Q_0^{ab..k}+\eta^{(ab}{\cal A}^{c..k)}.\label{eq:start}
\end{align}
\item Form subsequent traces with respect to $\eta_{ab}$, $\eta_{cd}$, etc., until only a vector or scalar expression remains.
\item Solve the final equation for ${\cal A}$ (if $k$ is even) or ${\cal A}^a$ (if $k$ is odd) in terms of $Q$ or $Q^a$.
\item Substitute the result into the preceding (rank-2 or rank-3) equation and solve for ${\cal A}^{ab}$ (for even $k$) or ${\cal A}^{abc}$ (for odd $k$).
\item Repeat iteratively until solutions for all the indexed ${\cal A}$ quantities are obtained in this manner.
\item Finally, substitute these solutions into (\ref{eq:start}), which is solved trivially for the trace-free $Q_0^{ab..k}$.
\end{enumerate}

It may be tempting to think that computing the trace-free representation of a rank-$k$ tensor can utilize results from the rank-$(k-1)$ or rank-$(k-2)$ computation. This, however, is not the case. At each rank, the computation relies on unique terms, despite the apparent, formal similarities between expressions.

We implemented this algorithm using the Maxima computer algebra system, as presented in \ref{app:A}. Although we have also obtained a closed form representation (see Sec.~\ref{sec:closedform} below) the iterative algorithm presented here may be more convenient in some situations. This implementation can compute the symbolic form of the decomposition for tensor ranks up to rank-8 on desktop computers with typical hardware and memory resources. (Higher rank tensors exhaust the memory allocation of the LISP language implementation underlying the Maxima language. The symbolic result in any case becomes too complex to be of practical use.)

This Maxima implementation of our trace-free decomposition algorithm can also be used either directly for numerical computation within the Maxima system, or it can serve as a model for numerical implementations. The Maxima session shown in \ref{app:B} demonstrates using this algorithm to explicitly compute matrix elements, in this case the rank-3 STF tensor characterizing the octupole gravitational potential of a point mass $m$ located at coordinates $(0,0,z)$ in a Cartesian coordinate system.

The trace-free nature of the result is readily seen. By assigning values to the parameters $m, z$ the final expression can be numerically evaluated. Specific cases, in principle up to arbitrary rank, in actual practice up to rank-8, can be just as readily evaluated. It is not necessary to display large, untractable interim expressions; a few additional lines of Maxima code can be used, as needed, to compute results numerically and then print only unique nonzero elements of the resulting tensor.

\section{Closed-form representation}
\label{sec:closedform}

Recurring patterns in this solution appear to suggest that a closed-form can be obtained. Indeed, such a representation has been derived for the case of $n=3$ dimensions in \cite{TrautmanPiraniBondi} for use in the study of gravitational radiation; see also \cite{1980RvMP...52..299T,Blanchet-Damour:1986,Damour1991}.

Following the approach in \cite{TrautmanPiraniBondi}, in the general $n$-dimensional case, we obtain by induction the expression
\begin{align}
Q_0^{a_1a_2..a_k}=&Q^{a_1a_2..a_k}+\sum_{p=1}^{\lfloor k/2\rfloor}(-1)^p\frac{\displaystyle{k!\eta^{(a_1a_2}..\eta^{a_{2p-1}a_{2p}}Q^{a_{2p+1}..a_k)}}}{\displaystyle{(k-2p)!2^p\prod_{i=1}^p[i n + 2i(k-i-1)] }}.
\end{align}
The product term in the denominator can, in turn, be expressed using gamma functions and simplified:
\begin{align}
\prod_{i=1}^p[i n + 2i(k-i-1)]=\frac{\Gamma(p+1)(-2)^p\Gamma(p+2-\frac{1}{2}n-k)}{\Gamma(2-\frac{1}{2}n-k)}=\frac{p! (n+2k-4)!!}{[n+2k-2(p+2)]!!},
\end{align}
leading to the final trace-free representation of the fully symmetric, $n$-dimensional, rank-$k$ tensor $Q^{a_1a_2..a_k}$ in the form
\begin{align}
Q_0^{a_1a_2..a_k}=Q^{a_1a_2..a_k}+\sum_{p=1}^{\lfloor k/2\rfloor}(-1)^p
\frac{k![n+2k-2(p+2)]!!}{\displaystyle{2^p p! (k-2p)! (n+2k-4)!!}}\eta^{(a_1a_2}..\eta^{a_{2p-1}a_{2p}}Q^{a_{2p+1}..a_k)}.
\label{eq:finalQ0}
\end{align}

This closed-form representation can also be readily implemented using computer algebra, as shown in \ref{app:STF-cf}. To demonstrate the practical applicability of our results, in \ref{app:STF-mom} we present the coordinate combinations needed to compute the STF mass moments of rank $k=5..8$ that are not readily available in the literature.

\section{Conclusions}
\label{sec:conclusions}

Symmetric trace-free tensors provide a versatile alternative to spherical harmonics in the investigation of many problems in physics, including the topic that inspired our investigation, gravitational lensing \cite{SGL2021a,SGL2021f,SGL2021g}. The primary advantage of the STF tensor representation is its coordinate independence. This allows us to work using the coordinate system of our choice, which includes the convenient Cartesian coordinate system. Nonetheless, the STF representation is less well known than that with the spherical harmonics. This is likely due to the fact that their use, especially the use of higher-rank STF tensors, can be computationally daunting, and that symbolic results suitable for analytical calculations are not readily available.

The STF representation often begins by obtaining a tensor, symmetric in all indices, but not yet trace-free. The corresponding STF tensor is obtained using the unique trace-free representation of this tensor. In principle, finding this representation requires solving a potentially large system of linear equations. The result that we developed relies on the unique properties of fully symmetric tensors to compute this decomposition efficiently in a coordinate-independent form. This may be done not only to arbitrary order, but also in the geometries with arbitrary dimensions.

Though the principles are well known, our result presents several advantages. First, it is valid in an arbitrary metric manifold with a nondegenerate metric, in arbitrary dimensions. Second, it is a closed-form expression: instead of relying on generic methods like matrix inversion or linear equation solvers, it relies on the unique properties of symmetric tensors. As a consequence, the results are readily presented in symbolic form, suitable for use in analytical expressions and for use in computer algebra.

As such, this method is usable in many applied contexts, including electromagnetism, geodesy and celestial mechanics, as well as the subject of our investigation, gravitational lensing. As the solution works in arbitrary dimensions, it may also find use in the development and application of new theories such as higher-dimensional theories of gravitation. Our results can also help solve wave propagation equations in many areas of wave physics.

In \ref{app:STF-mom} we present the STF coordinate combinations that are required to compute the STF mass multipole moments for rank $r=5..8$, which can be used to study gravitational lensing by complex mass distributions.  The work on using these results is ongoing; results, when available, will be published elsewhere.

\section*{Acknowledgments}

This work in part was performed at the Jet Propulsion Laboratory, California Institute of Technology, under a contract with the National Aeronautics and Space Administration.
VTT acknowledges the generous support of Plamen Vasilev and other Patreon patrons.

%

\bibliographystyle{ws-ijac}
\bibliography{trace}

\vskip 2em

\appendix

\section{Maxima implementation}
\label{app:A}

\verbatimfont{\footnotesize}%
\begin{verbatim}
/* makeQ0: Generate trace-free decomposition of a symmetric tensor.
 *
 * Usage: makeQ0(N, [d, g, Q]);
 *
 * parameters: N: rank of the tensor to be decomposed (integer, >1)
 *             d (optional): Number of dimensions (can be symbolic)
 *             g (optional): Name of the metric (default: n)
 *             Q (optional): Name of the tensor
 *
 * Given the rank-d contravariant tensor Q that is fully symmetric
 * in all indices, this program defines the itensor components of Q0,
 * its trace-free decomposition.
 *
 * Side effects: 1) loads itensor; 2) redefines imetric and dim,
 * 3) defines symmetry properties for the metric and Q, 4) defines
 * contraction properties for Q and Qa, and 5) defines the itensor
 * components for Qa and Q0.
 *
 * Example:
 *
 * (%i1) load("makeQ0.mac");
 * (%o1)                             makeQ0.mac
 * (%i2) makeQ0(2,3);
 * (%o2)                                done
 * (%i3) ishow(Q0([],[a,b]))$
 *                                            a b
 *                                   a b   Q n
 * (%t3)                            Q    - ------
 *                                           3
 * (%i4) makeQ0(3,n,g,T);
 * (%o4)                                done
 * (%i5) ishow(T0([],[a,b,c]))$
 *                       a b  c       a  b c      a c  b
 *                    2 g    T     2 T  g      2 g    T     a b c
 * (%t5)           (- ---------) - --------- - --------- + T
 *                     2 n + 4      2 n + 4     2 n + 4
 *
 * (c) 2021 Viktor T. Toth (https://www.vttoth.com/)
 *
 * This program is free software; you can redistribute it and/or
 * modify it under the terms of the GNU General Public License as
 * published by the Free Software Foundation; either version 2 of
 * the License, or (at your option) any later version.
 *
 * This program is distributed in the hope that it will be
 * useful, but WITHOUT ANY WARRANTY; without even the implied
 * warranty of MERCHANTABILITY or FITNESS FOR A PARTICULAR
 * PURPOSE.  See the GNU General Public License for more details.
 */

makeQ0(N,[L]):=block
(
 [AI,II,I3,E,i,j,a,q,q0],
 if not get('makeQ0,'version) then
 (
  put('makeQ0,'20210720,'version),
  load(itensor)
 ),
 if not integerp(N) or N < 2 then
   return("makeQ0: first argument must be an integer greater than 1!"),
 if length(L)>2 then (q:L[3],q0:concat(L[3],"0"),a:concat(L[3],"a"))
                else (q:Q,q0:Q0,a:Qa),
 if length(L)>1 then imetric(L[2]) else imetric(n),
 if length(L)>0 then dim:L[1] else dim:'d,
 decsym(imetric,2,0,[sym(all)],[]),
 decsym(imetric,0,2,[],[sym(all)]),
 defcon(a,a,a),
 defcon(q,q,q),
 for i from 2 thru N-2 do
 (
  decsym(a,i,0,[sym(all)],[]),
  decsym(a,0,i,[],[sym(all)]),
  decsym(q,i,0,[sym(all)],[]),
  decsym(q,0,i,[],[sym(all)])
 ),
 remcomps(a),
 AI:makelist(eval_string(concat("a",i)),i,1,N),
 II:makelist(eval_string(concat("i",i)),i,1,N),
 I3:makelist(eval_string(concat("i",i)),i,3,N),
 E:makelist(0,i,0,(N+1)/2),
 E[1]:(canform(contract(expand(kdels(II,AI)*ev(imetric)([],[i1,i2])*a([],I3))))),
 for i thru N-1 step 2 do (
  E[(i+3)/2]:(canform(contract(contract(expand(E[(i+1)/2]*
              ev(imetric)([AI[i],AI[i+1]]))))))
 ),
 if evenp(N) then components(a([],[]),rhs(solve(q([],[])=E[N/2+1],a([],[]))[1])),
 for i from N+(if evenp(N) then -1 else 0) thru 3 step -2 do
  components(a([],makelist(AI[j],j,i,N)),
             rhs(solve(q([],makelist(AI[j],j,i,N))=ev(E[(i+1)/2],a),
                       a([],makelist(AI[j],j,i,N)))[1])),
 components(q0([],AI),q([],AI)-ev(E[1],a))
)$
\end{verbatim}

\section{Maxima usage example}
\label{app:B}

The following Maxima program computes the rank-3 STF tensor corresponding to the octupole moment of the gravitational field of a point mass $m$ located at coordinates $(0,0,z)$ in a Cartesian coordinate system.

\verbatimfont{\footnotesize}%
\begin{verbatim}
(%i1) load("makeQ0.mac")$
(%i2) makeQ0(3,3,g,T)$
(%i3) components(T([],[c]),g([a,b],[])*T([],[a,b,c]))$
(%i4) EQ:T0([],[a,b,c])$
(%i5) EQ:EQ,T$
(%i6) EQ:subst(T0[a][b,c],'T0[a,b,c],ic_convert('T0([],[a,b,c])=EQ))$
(%i7) lg:ident(3)$
(%i8) ug:ident(3)$
(%i9) dim:3$
(%i10) T0:[0,0,0]$
(%i11) for i thru dim do T0[i]:zeromatrix(dim,dim)$
(%i12) for i thru 3 do for j thru 3 do for k thru 3 do T[i,j,k]:0$
(%i13) T[3,3,3]:m*z^3$
(%i14) ev(EQ)$
(%i15) T0:factor(T0);
                                                      [      3                 ]
        [                 3 ]  [ 0    0       0    ]  [   m z                  ]
        [              m z  ]  [                   ]  [ - ----    0       0    ]
        [   0     0  - ---- ]  [                 3 ]  [    5                   ]
        [               5   ]  [              m z  ]  [                        ]
        [                   ]  [ 0    0     - ---- ]  [              3         ]
(%o15) [[   0     0    0    ], [               5   ], [           m z          ]]
        [                   ]  [                   ]  [   0     - ----    0    ]
        [      3            ]  [         3         ]  [            5           ]
        [   m z             ]  [      m z          ]  [                        ]
        [ - ----  0    0    ]  [ 0  - ----    0    ]  [                      3 ]
        [    5              ]  [       5           ]  [                 2 m z  ]
                                                      [   0       0     ------ ]
                                                      [                   5    ]
\end{verbatim}

Lines 1--2 load the STF program and run it for the rank-3 3-dimensional case, using the symbols $g$ and $T$ for the metric tensor and the moment of inertia tensor, respectively.

Line 3 establishes the components of the traces of $T$ using the formalism of the maxima {\tt itensor} package.

Lines 4--6 set up an equation using the {\tt ic\_convert} facility of {\tt itensor}, which can convert an abstract index expression into a set of matrix element operations.

Lines 7--9 establish the metric and the dimensionality of the problem.

Lines 10--13 set up a rank-3 object as a vector of three matrices.

Finally, line 14 evaluates the {\tt ic\_convert} expression and line 15 displays the result.

\section{Closed form implementation}
\label{app:STF-cf}

The closed-form representation of the STF decomposition described in Sec.~\ref{sec:closedform} can be implemented using the following simple Maxima function:

\verbatimfont{\footnotesize}%
\begin{verbatim}
/* formQ0: Evaluate trace-free decomposition of a symmetric tensor.
 *
 * Usage: Q0(Q);
 *
 * parameters: Q: itensor expression
 *
 * Given the tensor Q that is fully symmetric in all indices, this
 * function returns the corresponding STF tensor.
 *
 * Side effects: None.
 *
 * Example:
 *
 * (%i1) load("formQ0.mac")$
 * (%i2) load(itensor)$
 * (%i3) dim:3$
 * (%i4) decsym(g,0,2,[],[sym(all)])$
 * (%i5) imetric(g)$
 * (%i6) flipflag:true$
 * (%i7) defcon(Q,Q,Q)$
 * (%i8) decsym(Q,0,3,[],[sym(all)])$
 * (%i9) ishow(canform('kdels([a,b,c],[j,k,l])*Q0(Q([],[a,b,c]))))$
 *                                          %1  %2 %3      j k l
 *                                       3 Q   g      kdels
 *              %1 %2 %3      j k l                        %1 %2 %3
 * (%t9)       Q         kdels         - --------------------------
 *                            %1 %2 %3               5
 *
 * (c) 2021 Viktor T. Toth (https://www.vttoth.com/)
 *
 * This program is free software; you can redistribute it and/or
 * modify it under the terms of the GNU General Public License as
 * published by the Free Software Foundation; either version 2 of
 * the License, or (at your option) any later version.
 *
 * This program is distributed in the hope that it will be
 * useful, but WITHOUT ANY WARRANTY; without even the implied
 * warranty of MERCHANTABILITY or FITNESS FOR A PARTICULAR
 * PURPOSE.  See the GNU General Public License for more details.
 */

Q0(_Q):=block
([_a,_k,_i,_j,_l,_g],
  _a:indices(_Q)[1],
  _k:length(_a),
  _g:ev(imetric),
  _Q+canform(contract(sum(
    (-1/2)^_p*(dim+2*_k-2*(_p+2))!!/_p!/(_k-2*_p)!/(dim+2*_k-4)!!
    *canform(contract(expand(
    kdels(makelist(concat('_i,_j),_j,1,_k),_a)
    *prod(_g([],[concat('_i,2*_i-1),
    concat('_i,2*_i)])*_g([concat('_j,2*_i-1),concat('_j,2*_i)]),_i,1,_p)
    *sublis(makelist(_a[_l]=(flatten([makelist(concat('_j,_i),_i,1,2*_p),
    makelist(concat('_i,_i),_i,2*_p+1,_k)]))[_l],_l,1,_k),_Q)))),
  _p,1,_k/2)))
)$
\end{verbatim}

Before running this subroutine, the {\tt itensor} package of Maxima must be loaded. The following usage example demonstrates the use of this formulation, used in conjunction with the powerful capabilities of {\tt itensor} to apply declared symmetries, and obtain the result in a particularly compact form:

\verbatimfont{\footnotesize}%
\begin{verbatim}
(%i1) load("formQ0.mac")$
(%i2) load(itensor)$
(%i3) dim:3$
(%i4) decsym(g,0,2,[],[sym(all)])$
(%i5) imetric(g)$
(%i6) flipflag:true$
(%i7) defcon(x,x,r2)$
(%i8) components(r2([],[]),r^2)$
(%i9) ishow(canform('kdels([a,b,c],[j,k,l])*Q0(x([],[a])*x([],[b])*x([],[c]))))$
                                         %1  %2 %3      j k l     2
                                      3 x   g      kdels         r
           %1  %2  %3      j k l                        %1 %2 %3
(%t9)     x   x   x   kdels         - -----------------------------
                           %1 %2 %3                 5
(%i10) ishow(canform('kdels([a,b,c,d],[j,k,l,m])
                     *Q0(x([],[a])*x([],[b])*x([],[c])*x([],[d]))))$
          %1 %2  %3 %4      j k l m      4
       3 g      g      kdels            r
                            %1 %2 %3 %4
(%t10) -----------------------------------
                       35
           %1  %2  %3 %4      j k l m      2
        6 x   x   g      kdels            r
                              %1 %2 %3 %4       %1  %2  %3  %4      j k l m
      - ------------------------------------ + x   x   x   x   kdels
                         7                                          %1 %2 %3 %4
\end{verbatim}

\section{STF multipole moments}
\label{app:STF-mom}

Despite their advantages compared to the method of spherical harmonics, STF multipole moments are not yet widely used modern physics (see discussion in \cite{SGL2021f}). This is primarily due to the fact that they are difficult to compute by hand. Using our results, especially with computer algebra, these tedious computations can be trivially simplified. Here we present an example, STF mass multipole moments of increasing rank.

We use the usual definition of the STF mass multipole moments \cite{Soffel-Han:2019}:
{}
\begin{eqnarray}
Q_0^{a..k}=\int d^3{\vec x} \, \rho({\vec x})\, x^{<a}..x^{k>},
\label{eq:Iab}
\end{eqnarray}
where angle brackets represent symmetrization and trace removal. The known coordinate combinations needed to compute the lowest Cartesian STF multipole moments \cite{Soffel-Han:2019} are given as
{}
\begin{align}
x^{<a} x^{b>}&{}=x^a x^b-\tfrac{1}{3}r^2\eta^{ab},\label{eq:stf2}\\
x^{<a}x^bx^{c>}&{}=x^a x^b x^c-\tfrac{1}{5}r^2\Big(\eta^{ab}x^c+\eta^{bc}x^a+\eta^{ca}x^b\Big)= x^a x^b x^c-\tfrac{3}{5}r^2\eta^{(ab}x^{c)},\label{eq:stf3}\\
x^{<a} x^bx^cx^{d>}&{}=x^ax^bx^cx^d-\tfrac{6}{7}r^2x^{(a}x^b\eta^{cd)}+\tfrac{3}{35}r^4\eta^{(ab}\eta^{cd)},\label{eq:stf4}
\end{align}
in precise correspondence with (\ref{eq:Q0-r2}), (\ref{eq:Q0-r3}) and (\ref{eq:Q0-r4}).

To our surprise, higher rank combinations, $k>4$, are not readily available in the literature. To demonstrate the utility of our result, we compute the coordinate combinations needed for STF moments of rank $k=5..8$ in three dimensions, $n=3$. Using the Maxima implementation of our closed form result presented in \ref{app:STF-mom}, we can easily compute all cases for $k=5..8$:
{}
\begin{align}
\hskip -0.125in
x^{<a} x^b x^c x^d x^{e>}&{}= x^a x^b x^c x^d x^e-\tfrac{10}{9}r^2\eta^{(ab}x^{c}x^{d}x^{e)}+
\tfrac{5}{21}r^4\eta^{(ab}\eta^{cd}x^{e)},
\label{eq:stf5}\\
\hskip -0.125in
x^{<a} x^bx^cx^dx^ex^{f>}&{}=x^ax^bx^cx^d x^e x^f-\tfrac{15}{11}r^2\eta^{(ab}x^cx^d x^e x^{f)}\nonumber\\
&{}+\tfrac{5}{11}r^4\eta^{(ab}\eta^{cd} x^e x^{f)}-\tfrac{5}{231}r^6\eta^{(ab}\eta^{cd} \eta^{ef)},\label{eq:stf6}\\
\hskip -0.125in
x^{<a}x^bx^cx^dx^ex^fx^{g>}&{}=x^ax^bx^cx^dx^ex^fx^g-\tfrac{21}{13}r^2\eta^{(ab}x^cx^dx^ex^fx^{g)}\nonumber\\
&{}+\tfrac{105}{143}r^4\eta^{(ab}\eta^{cd}x^ex^fx^{g)}-\tfrac{35}{429}r^6\eta^{(ab}\eta^{cd}\eta^{ef}x^{g)},\\
\hskip -0.125in
x^{<a}x^bx^cx^dx^ex^fx^gx^{h>}
&{}=x^ax^bx^cx^dx^ex^fx^gx^h-\tfrac{28}{15}r^2\eta^{(ab}x^cx^dx^ex^fx^gx^{h)}\nonumber\\
&{}+\tfrac{14}{13}r^4\eta^{(ab}\eta^{cd}x^ex^fx^gx^{h)}-\tfrac{28}{143}r^6\eta^{(ab}\eta^{cd}\eta^{ef}x^gx^{h)}\nonumber\\
&{}+\tfrac{7}{1287}r^8\eta^{(ab}\eta^{cd}\eta^{ef}\eta^{gh)}.
\end{align}

These expressions may now be used to compute the STF moments for $k=5..8$, as we did in \cite{SGL2021g}.

\end{document}